\catcode`\@=11
{\count255=\time\divide\count255 by 60 \xdef\hourmin{\number\count255}
        \multiply\count255 by-60\advance\count255 by\time
   \xdef\hourmin{\hourmin:\ifnum\count255<10 0\fi\the\count255}}
\def\ps@draft{\let\@mkboth\@gobbletwo
    \def\@oddhead{}
    \def\@oddfoot
       {\hbox to 7 cm{$\scriptstyle Draft\ version:\ \draftdate$
       \hfil}
       \hskip -7cm\hfil\rm\thepage \hfil}
    \def\@evenhead{}\let\@evenfoot\@oddfoot}
\catcode`\@=12
\def\draftdate{\number\month/\number\day/\number\year\ \ \ \hourmin }

\documentstyle[12pt]{article}
\newcommand{\be}{\begin{eqnarray}}
\newcommand{\en}{\end{eqnarray}\vs 0.5 cm}
\newcommand{\non}{\nonumber}
\newcommand{\no}{\noindent}
\newcommand{\vs}{\vskip}
\newcommand{\hs}{\hspace}

\newcommand{\p}{\partial}

\begin{document}
\
\begin{center}
\no ON QUANTUM GROUP SYMMETRIES OF CONFORMAL FIELD
THEORIES\hs{0.02 cm}\footnote{\hs{-0.02 cm}$^)$\hs{0.04 cm}talk given
at the XX$^{\rm th}$ International
Conference on Differential Geometric Methods in Theoretical Physics,
New York,
June 3-7, 1991}$\hs{-0.04 cm}^)$

\vskip 1.2 cm

\no FERNANDO FALCETO
\vskip 0.1 cm

\no{\it Department of Mathematics, Rutgers University, New Brunswick
08903 NJ, USA}
\vs 0.5 cm

\no and
\vs 0.5 cm

\no KRZYSZTOF GAW\c{E}DZKI
\vs 0.1 cm

\no {\it Institut des Hautes Etudes Scientifiques, 91440 Bures-sur-Yvette,
France}
\vs 1 cm
\end{center}

\begin{abstract}

The appearance of quantum groups in conformal field theories is traced back
to the Poisson-Lie symmetries of the classical chiral theory.
A geometric quantization of the classical theory deforms the Poisson-Lie
symmetries to the quantum group ones. This elucidates the fundamental role
of chiral symmetries that quantum groups play in conformal models.
As a byproduct, one obtains a more geometric approach to the representation
theory of quantum groups.

\end{abstract}


\vskip 1 cm

\no{\bf 1.\ \ Canonical structure of the chiral WZW models}
\vs 0.4 cm

Quantum Groups$^{1-3}$ (QGs) have entered into Conformal Field
Theory (CFT) through
the back door: it was discovered that the exchange properties of (some)
CFT chiral
vertex operators lead to the braid group representations related to
QGs$^{4,5}$, that the QG $6j$ symbols may be realized as
braiding matrices of
those operators$^{6-8}$ and that the CFT fusion rules are related
to the tensor
product decomposition of quantum group representations at roots of
unity$^{6,9}$. In view of these relations it was
becoming clear that QGs should
play a role of new symmetries of chiral CFTs$^{10-19}$. Since symmetries
play a fundamental role in physics, it would be desirable to have an
approach
to CFTs which puts QG symmetries in the foreground. This is a report about
an attempt at such an approach. The main idea we follow is to start at the
classical level and to identify classical symmetries of the chiral CFTs
which
upon quantization become QG symmetries. This idea was pursued before in a
series of papers started by Ref.$^{20}$, or in the more direct sense, by
the St. Petersburg school in Refs.$^{21,22}$, see also Ref.$^{23,24}$.
The present exposition is
based on a paper in preparation extending the results of Ref.$^{25}$.
Another (possibly more fundamental) approach has been proposed recently
in Refs.$^{26-28}$ were an attempt was made to
build a lattice version of CFT's with
local QG symmetry whose global part survives the continuum limit.
Although different from ours, this approach also has some common points
with our work.
\vs 0.2 cm

We shall concentrate below on the Wess-Zumino-Witten (WZW) model$^{29}$ of
CFT although our analysis may be easily extended to various coset theories.
In that model the classical field configurations are given by Lie group
$G$-valued functions $g(x^0,x^1)$ of two-dimensional Minkowski space-time
coordinates satisfying

\be
\p_-(g\p_+g^{-1})=0
\en

\no where $\p_\pm = (\p_1\pm\p_0)$. On the cylinder
$Z=\{(x^0,x^1\ {\rm mod}\hs{0.07 cm}2\pi)\}$, the classical solutions are
of the form

\be
g(x^0,x^1)=g_L(x^+)\hs {0.05 cm}g_R(x^-)^{-1}
\en

\no where $x^\pm=x^1\pm x^0$ and

\be
g_{L,R}(x+2\pi)=g_{L,R}(x)\gamma_{L,R}\hs{0.5 cm}
\en

\no with $\gamma_L=\gamma_R\in G$ (so that $g$ is periodic in $x^1$).
Let us denote by $P$ the space
of all solutions of Eq.\hs{0.08 cm}1 on $Z$, i.e. the phase space of
the complete WZW model.  Introducing the phase spaces
$P_L$ ($P_R$) of left- (right-)movers as the spaces
of smooth maps $g_L$ ($g_R$) satisfying relation 3 and
denoting by $\Delta$ the subset of $P_L\times P_R$ with equal monodromies
$\gamma_L=\gamma_R$, we have

\be
P\ =\ \Delta/G
\en

\no where $G$ acts by

\be
(g_L,g_R)\longmapsto(g_Lg_0,g_Rg_0)
\en

\no and describes the ambiguity of representation 2. (Notice that
under transformation 5, $\gamma_{L,R}\longmapsto Ad_{g_0^{-1}}
(\gamma_{L,R})$).
The canonical structure of the WZW theory is described by the (uniquely
determined) symplectic form $\Omega$ on phase space $P$ which may be
written$^{25}$ as $\Omega_L-\Omega_R$
where

\be
\Omega_{L,R}\ =\ (4\pi)^{-1}k\int\limits_0^{2\pi}{\rm tr}(g_{L,R}^{-1}
dg_{L,R})
\wedge\p_x(g_{L,R}^{-1}dg_{L,R})\hs{1.5 cm}\non\\
+\ (4\pi)^{-1}k\ {\rm tr}\hs{0.06 cm}({g_{L,R}}^{-1}
dg_{L,R})(0)\wedge(d\gamma_{L,R})\gamma_{L,R}^{-1}\
-\ (4\pi)^{-1}k\ \rho(\gamma_{L,R})\hs{0.2 cm}
\en

\no are 2-forms on $P_{L,R}$. In Eq.\hs{0.08 cm}6, $\rho$ is
an arbitrary 2-form on $G$. Such (and only such) an ambiguity arises
because only the restriction of $\Omega_L-\Omega_R$ to $\Delta\subset
P_L\times
P_R$ enters in the determination of (unique) $\Omega$.
\vs 0.3 cm

$\Omega_L$ seems to be a natural candidate for the 2-form defining the
canonical structure for the left-movers (and similarly for the
right-movers).
Somewhat surprisingly, however, there are problems with such an
interpretation. First, $\Omega_L$ is not unique. Much worse, a
straightforward
calculation shows that

\be
d\Omega_L(g_L)\ =\ (12\pi)^{-1}k\ {\rm tr}\hs{0.05 cm}(\gamma_L^{-1}
d\gamma_L)^{\wedge 3}\ -
\ (4\pi)^{-1}k\ d\rho(\gamma_L)
\en

\no so that $d\Omega_L$ can never be zero globally as ${\rm tr}
\hs{0.05 cm}
(\gamma^{-1}d\gamma)^3$ is not an exact form on (simple) $G$.
At least three possible ways out of the latter difficulty may be
considered:
\vs 0.1 cm

\no 1. One may use ambiguity 5 to restrict $P_L$ to maps with specific
monodromies. For example, for compact $G$ we may introduce $P^{res}_L$
corresponding to monodromies $\gamma_L$ in the Cartan subgroup
$T\subset G$.
Similar choices may be made for non-compact groups.
On $P^{res}_L$, $d\Omega_L=0$ whenever $d\rho=0$ on $T$. This is
an approach parallel to that of Refs.$^{12,20,21,30-32}$
which worked with the diagonal monodromy.
\vs 0.1 cm

\no 2. By choosing suitable $\rho$ with $d\rho={\rm tr}\hs{0.05 cm}
(\gamma^{-1}d\gamma)^{\wedge 3}/3$ on an open
dense subset in $G$, we may obtain
a singular symplectic structure on $P_L$ which leads however to a regular
Poisson bracket (at the singularities of the symplectic form the Poisson
structure ceases to be non-degenerate). We shall pursue this approach here.
\vs 0.1 cm

\no 3. One may interpret Eq.\hs{0.08 cm}7 as an obstruction to
closeness of $\Omega_L$
which would be reflected in the violation of the Jacobi identity for
the Poisson bracket on $P_L$. This leads to the appearance of
classical counterparts of Drinfeld's quasi-Hopf algebras$^{33}$.
We shall discuss this approach elsewhere. See also Ref.$^{34}$.
\vs 0.2 cm

The Poisson bracket induced by $\Omega_L$ on $P_L$ has the form

\be
\{g_L(x)_1,g_L(y)_2\}\ =\ -2\pi k^{-1}\ g_L(x)_1\hs{0.07 cm}
g_L(y)_2\ r^{\pm}(\gamma_L)
\en

\no in a shorthand notation where $g_L(x)_1=g_L(x)\otimes 1$,
$g_L(y)_2=1\otimes g_L(y)$ and $r^{\pm}(\gamma_L)\in {\cal G}^{\bf C}
\otimes{\cal G}^{\bf C}$ with ${\cal G}$ the Lie algebra of $G$ are all
four
treated as endomorphisms of $V_1\otimes V_2$
with $V_i$ representation spaces of $G$. $\pm$ sign in $r^{\pm}$
is used depending on whether $x<y$ or $x>y$.

Poisson bracket 8 has, in general, one nasty feature:
non-locality.
The right hand side depends not only on the values of $g_L$ at points
appearing
on the left hand side but also on the non-local monodromy of $g_L$.
It is then rather natural to ask if for certain choices of (singular)
2-form $\rho$ the monodromy dependence of $r^{\pm}$ disappears and
Poisson bracket 8 becomes local\footnote{$^{)}$\hs{0.07 cm}the chiral
notion of locality should
not be confused with physical locality in the complete theory holding
in any case}$^{)}$. Those choices
would necessarily lead to matrices $r^{\pm}$ satisfying the Classical
Yang-Baxter Equation (CYBE) (without spectral parameter)

\be
[r^\pm_{12},r^\pm_{13}]+[r^\pm_{12},r^\pm_{23}]
+[r^\pm_{13},r^\pm_{23}]\ =\ 0
\en

\no equivalent to the Jacobi identity for bracket 8 with
constant $r^{\pm}$. (Eq.\hs{0.08 cm}9 should be read as an
equality between endomorphisms of $V_1\otimes V_2\otimes V_3$).
Moreover,

\be
r^{\pm}=r\pm\kappa
\en

\no where $r\in{\cal G}^{\wedge 2}$ and $\kappa=\sum t^a\otimes t^a$
is the quadratic Casimir of $G$ (with the generators $t^a$ normalized by
${\rm tr}\hs{0.05 cm}t^at^b=\delta^{ab}/2$).
\vs 0.2 cm

Conversely, suppose that we are given a pair $r^{\pm}$ satisfying
relations 9 and 10. Consider, following Ref.$^{35}$, the subspace

\be
{\cal G}_r\ \equiv\ \{(r^+\rfloor v,r^-\rfloor v)\ |
\ v\in{\cal G}^{*{\bf C}}\}\ \subset\ {\cal G}^{\bf C}\oplus
{\cal G}^{\bf C}\ .
\en

\no Eqs.\hs{0.01 cm} 9 and 10 imply that ${\cal G}_r$ is a complex Lie
subalgebra
of ${\cal G}^{\bf C}\oplus{\cal G}^{\bf C}$, see Ref.$^{35}$. Denote
by $G_r$ the
corresponding Lie group $\subset G^{\bf C}\times G^{\bf C}$ and
by $\iota$ the
map

\be
G^{\bf C}\times G^{\bf C}\ni(\gamma_+,\gamma_-)\ \longmapsto\ \gamma_+
\gamma_-^{-1}\in G^{\bf C}\ .
\en

\no The restriction of $\iota$ to $G_r$ is a covering map onto an open
dense subset $G^{\bf C}_0$ in $G^{\bf C}$.
Consider a (complex) 2-form $\rho$ on $G^{\bf C}_0$ defined in terms of
(multivalued) coordinates $(\gamma_+,\gamma_-)$ by

\be
\rho\ =\ {\rm tr}\hs{0.08 cm}\gamma_+^{-1}d\gamma_+\hs{0.05 cm}\wedge
\hs{0.05 cm}\gamma_-d\gamma_-^{-1}\ .
\en

\no The corresponding (singular) 2-form $\Omega_L$ leads to Poisson
bracket 8 which reproduces for $|x-y|<2\pi$ the constant matrices
$r^{\pm}$ on the right hand side.
\vs 0.3 cm

Summarizing: there is an ambiguity in defining canonical structure
of the chiral WZW theory. Possible local solutions are in one to one
correspondence with pairs $r^{\pm}$ of solutions of the CYBE.
\vs 0.3 cm

The best known example of a pair $r^{\pm}$ of solutions of the QYBE is
obtained
by taking in Eq.\hs{0.08 cm}10

\be
r\ =\ \sum\limits_{\alpha>0}(e_{\alpha}\otimes e_{-\alpha}-
e_{-\alpha}\otimes e_{\alpha})/2
\en

\no where the sum runs over the positive roots of $\cal G$
and $e_{\pm\alpha}$
are the corresponding nilpotent generators of ${\cal G}^{\bf C}$
normalized so that ${\rm tr}\hs{0.07 cm}e_{\alpha}e_{-\alpha}=1$.
The classification of solutions of the CYBE (without spectral parameter)
may be found in Ref.$^{36}$.
\vs 0.8 cm

\no{\bf 2. Poisson-Lie symmetries}.
\vs 0.4 cm

Phase space $P_L$ of the chiral WZW theory together with
Poisson bracket 8 with monodromy independent $r^{\pm}$ provide
an (infinite-dimensional) example of a Poisson manifold$^{37}$, i.e.
a manifold supplied with a field of 2-covectors whose contraction
with differentials of two functions on the manifold gives their Poisson
bracket\footnote{$^{)}$\hs{0.07 cm}in fact the example is not
quite conventional since
the Poisson structure on $P_L$ is complex}$^{)}$. In our case, the
Poisson
structure of $P_L$ comes from the inversion of a (complex) singular
symplectic
form $\Omega_L$. The notion of a Poisson manifolds is more general then
that
of a symplectic manifold. More importantly, it allows a natural
generalization of the notion of symmetry. Conventionally, we would say
that
$\Gamma$ is a symmetry group of the Poisson manifold $\Pi$ if $\Gamma$
acts
(from left or right) on $\Pi$ preserving its Poisson structure. The
generalized (Poisson-Lie) symmetries involve the notion of a
Poisson-Lie (PL)
group i.e. a Lie group $\Gamma$ provided with a Poisson structure
compatible
with the group multiplication$^{38,35}$. As for Lie groups, there is a
corresponding
infinitesimal notion: that of a bialgebra i.e. of a Lie
algebra $\Upsilon$
(of $\Gamma$)
together with a Lie algebra structure on the dual space $\Upsilon^*$,
both
compatible in a suitable way. For each bialgebra,
there is a dual bialgebra with
the roles of $\Upsilon$ and $\Upsilon^*$ interchanged. This duality lifts
to the (simply connected) PL groups which come in
pairs $(\Gamma$, $\Gamma^*)$.
\vs 0.2 cm

The simplest example of a PL group may be obtained by taking a Lie group
$\Gamma$ with the vanishing Poisson structure. The corresponding
Lie algebra
$\Upsilon$ becomes a bialgebra with the vanishing Lie bracket
on $\Upsilon^*$
and the corresponding dual PL group is $\Upsilon^*$ with addition as
the group
operation and with the Poisson bracket which to the linear functions on
$\Upsilon^*$ given by elements $\tau,\sigma\in\Upsilon$ assigns the
linear
function
given by $[\tau,\sigma]$. The symplectic leaves of $\Upsilon^*$ with
this Poisson
structure (i.e. connected components of common level sets of functions
on $\Upsilon^*$ with vanishing Poisson brackets with everybody else)
are exactly the coadjoint orbits in $\Upsilon^*$. As is well known, for
large class of Lie groups (e.g. for the compact ones), the coadjoint
orbits
are related to irreducible representations of the group$^{39}$.
\vs 0.2 cm

Another, less trivial example of a PL group is obtained by defining,
following Sklyanin$^{40}$, a Poisson structure on a (complex)
Lie group $\Gamma$ by putting

\be
\{\gamma_1,\gamma_2\}_{_{\rm Skl}}\ =\
2\pi k^{-1}\hs{0.05 cm}[\hs{0.08 cm}\gamma_1\hs{0.06 cm}
\gamma_2\hs{0.08 cm},
\hs{0.08 cm}r^\pm\hs{0.08 cm}]
\en

\no in the notation of Eq.\hs{0.08 cm}8 ($\gamma$ is the matrix
function on $\Gamma$
given by a representation, $\gamma_1=\gamma\otimes 1$ etc.; both signs
give the same Poisson bracket).
The Lie algebra of the dual PL group $\Gamma^*$ may be identified with
$\Upsilon_r$ defined as in 11 via the map $v\longmapsto(r^+\rfloor v,
r^-\rfloor v)$ and $\Gamma^*$ itself with $\Gamma_r\subset\Gamma
\times\Gamma$.
The symplectic leaves of $\Gamma^*$ become then connected components
of the
preimages of the conjugacy classes in $\Gamma$ under the covering
map $\iota$,
see Ref.$^{35}$. They play a role in the representation theory of
quantum groups.
\vs 0.2 cm

We shall say that $\Gamma$ is a PL symmetry of Poisson manifold $\Pi$
if it is a PL group which acts on $\Pi$ so that the corresponding map
$\Pi\times\Gamma\longrightarrow\Pi$ is Poisson i.e. preserves the
Poisson brackets. In the case of $\Gamma$ with the vanishing Poisson
structure this definition is equivalent to demanding that the action
of $\Gamma$ preserves the Poisson structure of $\Pi$ so that the
notion of
a PL symmetry generalizes that of a standard (Lie) symmetry.
\vs 0.2 cm

$P_L$ with the Poisson structure that we have introduced has several
symmetries. First, there are conventional symmetries:
\vs 0.1 cm

\no 1. Loop group symmetry. $LG$ (the group of periodic maps $h(x)$ with
values in $G$) acts on $P_L$ by $g_L\longmapsto hg_L$
preserving $\Omega_L$
and the corresponding Poisson structure.
\vs 0.1 cm

\no 2. Conformal symmetry. Group $Diff_+(S^1)$ of orientation-preserving
diffeomorphisms $D$ of the circle acts by $g_L\longmapsto g_L\circ D$
again preserving $\Omega_L$.
\vs 0.2 cm

\no But $P_L$ also has $G_{\rm Skl}$ as a PL symmetry. Namely, $G$ acts
on $P_L$ by

\be
(g_L,g_0)\longmapsto g_Lg_0\ ,
\en

\no and map 16 preserves the Poisson brackets if $G$
is taken with Sklyanin bracket 15 (for real $G$ this defines a
complex Poisson structure on $G$ and the notion of a PL group should
be extended accordingly).
\vs 0.02 cm

Let us return to the general discussion of symmetries. If $\Gamma$
is a Lie symmetry of a symplectic manifold $\Pi$ one may often encode
its action in the so called moment map$^{41}$

\be
m: \Pi\ \longrightarrow\ \Upsilon^*
\en

\no such that if $\tau$ is in the Lie algebra $\Upsilon$
then the contraction of
$m$ with $\tau$ gives a hamiltonian function on $\Pi$ generating the
infinitesimal action of $\tau$. One also demands that the hamiltonian
of $[\tau,\sigma]$ be the Poisson bracket of hamiltonians
of $\tau$ and $\sigma$ or
in other words that $m$ be a Poisson map if $\Upsilon^*$ is taken with
the Poisson structure making it the dual PL group to $\Gamma$ with
the vanishing Poisson bracket (recall the discussion above).
An example at hand is the Sugawara energy-momentum tensor
$T=(2k)^{-1}{\rm tr}\hs{0.06 cm}J^2$ where the current
$J=ik(\p_xg_L)g_L^{-1}$. The quadratic differential $T$ may be
viewed as a
map

\be
P_L\ \longrightarrow\ Vect(S^1)^*
\en

\no into the dual of the space of vector fields on the circle
and is the moment map for the action of $Diff_+(S^1)$ on $P_L$.
\vs 0.2 cm

In some situations there are obstructions to existence of the moment
maps
as defined above$^{41}$. For example, current $J$ could be
viewed as a map of $P_L$
into $L{\cal G}^*$, the dual space to the Lie algebra of $LG$
but as such would not provide a Poisson map because of the central term
in the Poisson bracket of currents. Instead, one should consider
a central extension $\hat{L{\cal G}}\longrightarrow L{\cal G}$
of the loop algebra and treat $J$ as taking values
in $\hat{L{\cal G}}^*$
which leads to the following diagram of the Poisson maps:

\be
P_L\ \longrightarrow\ \hat{L{\cal G}}^*\ \longleftarrow\ L{\cal G}^*\ .
\en
\vs 0.1 cm

The notion of a moment map extends to the case of PL symmetries$^{42}$
where a moment map becomes an appropriate
Poisson map\footnote{$^{)}$\hs{0.07 cm}we imply here a slightly
more restrictive notion of a moment map then the one
of ``momentum mapping'' defined in Ref.$^{42}$}$^{)}$

\be
\Pi\ \longrightarrow\ \Gamma^*\ .
\en

\no Again, there might exist obstruction to the existence of moment maps
in strict sense. An example is provided by the case of PL symmetry of
$P_L$ considered above. Instead of a map like 20, we find here
a diagram of Poisson maps

\be
P_L\ \longrightarrow\ G\subset G^{\bf C}\ \longleftarrow\ G_r
\equiv (G^{\bf C})^*\ .
\en

\no Above, the leftmost arrow is the map $g_L\longmapsto\gamma_L(=
g_L(0)^{-1}g(2\pi))$ and the right one is $\iota$ of 12.
The Poisson structure on $G^{\bf C}$ is given by

\be
\{\hs{0.07 cm}\gamma_1\hs{0.07 cm},\hs{0.07 cm}\gamma_2
\hs{0.07 cm}\}\ =
\ -2\pi k^{-1}\hs{0.04 cm}(\hs{0.07 cm}r^{\pm}\hs{0.03 cm}
\gamma_1\hs{0.03 cm}
\gamma_2\hs{0.07 cm}-\hs{0.07 cm}\gamma_1\hs{0.03 cm}r^-\hs{0.03 cm}
\gamma_2
\hs{0.07 cm}
-\hs{0.07 cm}\gamma_2\hs{0.03 cm}r^+\hs{0.03 cm}\gamma_1\hs{0.07 cm}+
\hs{0.07 cm}\gamma_1\hs{0.03 cm}\gamma_2\hs{0.03 cm}r^{\mp}
\hs{0.07 cm})
\en

\no and the one on $G_r$ by

\be
\{\hs{0.07 cm}\gamma_{+1}\hs{0.03 cm},\hs{0.03 cm}\gamma_{+2}
\hs{0.03 cm}\}
\ =\ 2\pi k^{-1}\hs{0.03 cm}[\hs{0.07 cm}r^{\pm}\hs{0.03 cm},
\hs{0.03 cm}
\gamma_{+1}\hs{0.03 cm}\gamma_{+2}\hs{0.03 cm}]\ ,\non\\
\{\hs{0.07 cm}\gamma_{+1}\hs{0.03 cm},\hs{0.03 cm}\gamma_{-2}
\hs{0.03 cm}\}
\ =\ 2\pi k^{-1}\hs{0.03 cm}[\hs{0.07 cm}r^{+}\hs{0.03 cm},
\hs{0.03 cm}
\gamma_{+1}\hs{0.03 cm}\gamma_{-2}\hs{0.03 cm}]\ ,\\
\{\hs{0.07 cm}\gamma_{-1}\hs{0.03 cm},\hs{0.03 cm}
\gamma_{-2}\hs{0.03 cm}\}
\ =\ 2\pi k^{-1}\hs{0.03 cm}[\hs{0.07 cm}r^{\pm}\hs{0.03 cm},
\hs{0.03 cm}
\gamma_{-1}\hs{0.03 cm}\gamma_{-2}\hs{0.03 cm}]\ .\hs{0.05 cm}\non
\en

\no As we see, the monodromy plays the role of the (generalized)
moment map for the PL symmetry of the chiral phase space $P_L$.
\vs 0.8 cm

\no{\bf 3.\ \ Classical vertex-IRF transformation}
\vs 0.4 cm

Let us assume for concreteness that G is a simple compact group.
It is convenient to parametrize $g_L\in P_L$ writing

\be
g_L(x)\ =\ h(x)\hs{0.05 cm}{\rm e}^{ix\tau}\hs{0.05 cm}g_0^{-1}
\en

\no where $h\in LG$, $\tau$ is in the Cartan subalgebra ${\cal T}
\subset
{\cal G}$ and $g_0\in G$. This parametrization is not unique.
First, $\tau$
is determined up to the action of the affine Weyl group.
We may fix this ambiguity by taking $\tau$ from
the positive Weyl alcove $A\subset{\cal T}$. This will leave us only
with the possibility to multiply $h$ and $g_0$ on the right
by the same element of the Cartan subgroup $T$. In parametrization 24,

\be
\Omega_L(g_L)\ =\ \Omega_{L1}(h,\tau)\ +\ \Omega_{L2}(g_0,\tau)
\en

\no where

\be
\Omega_{L1}(h,\tau)\ =\ (4\pi)^{-1}k\int\limits_0^{2\pi}
{\rm tr}\hs{0.08 cm}[(h^{-1}dh)\wedge\p_x(h^{-1}dh)+2i
\tau(h^{-1}dh)^{\wedge 2}\non\\
-2i(d\tau)\wedge(h^{-1}dh)]
\en

\no and

\be
\hs{0.5 cm}\Omega_{L2}(g_0,\tau)\ =\ ki\hs{0.08 cm}{\rm tr}\hs{0.08 cm}
(d\tau)\wedge g_0^{-1}dg_0
+(4\pi)^{-1}k\hs{0.05 cm}{\rm tr}\hs{0.06 cm}g_0^{-1}dg_0
\wedge Ad_{{\rm e}^{2\pi i\tau}}(g_0^{-1}dg_0)\non\\\non
\en
\vs -1.9 cm
\be
- (4\pi)^{-1}k\hs{0.07 cm}\rho
(g_0{\rm e}^{2\pi i\tau}g_0^{-1})\ .\hs{-8.9 cm}
\en
\vs 0.1 cm

Symplectic form $\Omega_{L1}(h,\tau)$ is equal
to $\Omega_L|_{\rho\equiv0}$
of Eq.\hs{0.08 cm}6 restricted to the set of $g_L$ with monodromy
in the Cartan subgroup $T$, i.e. to $P^{res}_L$ introduced above.
$P^{res}_L$ plays the role of what has been called in Ref.$^{43}$ a
``model space'' for the Kac-Moody group $\hat{LG}$. It is a
symplectic space
which contains each (generic) coadjoint orbit of $\hat{LG}$ once.
Indeed,
if we fix $\tau$ in the Weyl alcove $A$, $\Omega_{L1}$ becomes a
(degenerate)
2-form on $LG$ which coincides with the pullback of the symplectic
form
of the coadjoint orbit of $\hat{LG}$ labeled
by $\tau$ by the natural map of $LG$ onto the orbit, see Ref.$^{44}$.
\vs 0.2 cm

Similarly, $\Omega_{L2}(g_0,\tau)$ may be viewed as the symplectic
form on the ``model space'' for the PL group $G$ with the Sklyanin
Poisson
structure. As was suggested above, for PL groups we should rather
talk about
the symplectic leaves of the dual group than about coadjoint orbits.
For $G$
(or $G^{\bf C}$) with the Sklyanin Poisson structure, the dual
group is
isomorphic to $G_r$ which covers by $\iota$ (an open dense subset of)
$G^{\bf C}$. Moreover, the symplectic leaves of the dual group
correspond
by $\iota$ to the conjugacy classes in $G^{\bf C}$.
Restricting to the compact group $G$ and its conjugacy classes
$\equiv\{g_0{\rm e}^{2\pi i\tau}g_0^{-1}\ |\ g_0\in G\}$ one may show
that,
in terms of $g_0$, the symplectic form of the symplectic leaves
coincides
with $\Omega_{L2}$ at fixed $\tau$.
\vs 0.2 cm

As we see, the chiral phase space $P_L$ may be realized as the
fibered product

\be
M_{KM}\times_AM_{PL}
\en

\no of the Kac-Moody  and the Poisson-Lie
model spaces over the Weyl alcove $A$, as summarized by
Eq.\hs{0.08 cm}24
\hs{-0.08 cm}\footnote{$^{)}$\hs{0.07 cm}more exactly
this is true for the open
dense set of $P_L$
obtained by exclusion of $\tau$ in $\p A$}$^{)}$.
The Poisson bracket of fields $g_L^{res}(x)\equiv h(x)\hs{0.05 cm}
{\rm e}^{ix\tau}$ on $P_L^{res}$ has also
form 8 but with $\tau$-dependent $r^\pm$, the classical counterparts
of the quantum group $6j$ symbols$^{21,22}$. Eq.\hs{0.05 cm}24
establishes
a relation
between those fields and fields $g_L$ with
monodromy-independent $r^\pm$
Poisson brackets. This is the classical version of the vertex-IRF
transformation for the (WZW) CFTs$^{10,11,16,17}$. It has similar
flavor as
field transformations described in Refs.$^{20,12,21,31,32}$ with the
important difference
that there the vertex versions of the fields still live on the phase
space with diagonal monodromy whereas our $g_L$'s are functionals
on the bigger phase space $P_L$ with general monodromy. As a result,
contrary
to Refs.$^{21,31,32}$, we may obtain vertex fields with arbitrary
solution
$r^\pm$ of the CYBE in the Poisson bracket. In particular, the
standard
r-matrix 14 may be used for any $G$ whereas in Refs.$^{31,32}$, for
$SU(N)$ with $N\ge 3$, different solutions were obtained.
\vs 0.8 cm

\no{\bf 4.\ \ Quantization}
\vs 0.4 cm

Let us briefly discuss how the preceding analysis may be extended
to the quantum theory. We shall give a more complete account in a
future publication.
A good idea is to use geometric quantization$^{45,46}$ which keeps
track of the
classical geometry. In view of presentation 28 of the chiral
phase space, we may first quantize model spaces $M_{KM}$ and $M_{PL}$
separately and then impose condition $\tau_{KM}=\tau_{PL}$ in the
quantum
theory.
\vs 0.2 cm

The geometric quantization of the Kac-Moody model space is more
or less standard. One takes the complex line bundle ${\cal L}_{KM}$
over $M_{KM}$ with the hermitian
connection of curvature $\Omega_{L1}$ (this is possible for $k$ an
integer)
and polarization ${\cal P}_{KM}$ of $M_{KM}$ given by the (complex)
tangent vectors annihilated by forms $d\tau$ and

\be
\int\limits_0^{2\pi}h^{-1}dh(x)\hs{0.06 cm}{\rm e}^{inx}dx\ ,\ \ \
{\rm tr}\hs{0.12 cm}e_\alpha\int\limits_0^{2\pi}h^{-1}dh(x)
\hs{0.05 cm}
dx\ ,\hs{0.4 cm}{\rm for}\ \ n<0\ ,\ \ \alpha<0\ .\non
\en

\no The space of quantum states is the homology of the sheaf of
${\cal P}_{KM}$-horizontal sections of ${\cal L}_{KM}$ (only $H^1$
contributes)
or, equivalently, of distributional ${\cal P}_{KM}$-horizontal
sections
of ${\cal L}_{KM}$. The latter are supported by $k\tau$ in the weight
lattice\footnote{$^{)}$\hs{0.07 cm}$k\tau$ is canonically
conjugate to the Cartan subgroup
component of the zero mode of $h$}$^{)}$.
For $\tau$ fixed at such a value, the problem reduces to the
geometric quantization of the corresponding coadjoint
orbit of $\hat{LG}$
isomorphic to $LG/T$ on which the polarization induces the standard
complex
structure$^{47}$. Over $LG/T$, ${\cal L}_{KM}$ becomes
a holomorphic line bundle
and the quantum states its holomorphic sections. This way, for
fixed $\tau$,
we recover the Borel-Weil construction of the irreducible
representation
space $\hat{V}_{k,\lambda}$ of $\hat{LG}$ corresponding to the
highest weight
$\lambda=k\tau$ and level $k$, see Ref.$^{47}$. It is still
better to use the improved
geometric quantization where states are half-densities$^{45}$ in
which case,
if we replace original $k$ by $k+h\hs{0.04 cm}{\check{}}$
where $h\hs{0.04 cm}
{\check{}}$ is the
dual Coxeter number of $G$, we end up, for fixed weight $\tau$, with
$\hat{V}_{k,\lambda}$ for $\lambda+\rho=(k+h\hs{0.04 cm}
{\check{}}\hs{0.04 cm})\tau$ where here $\rho$
denotes the Weyl vector $\sum\limits_{\alpha>0}\alpha/2$. In any case
the total spaces of states corresponding to $M_{KM}$ is

\be
V_{KM}\ =\ \bigoplus\limits_{{\rm integrable}\ \lambda}
\hat{V}_{k,\lambda}
\en

\no where integrable weights satisfy $(\lambda+\rho)/(k+h\hs{0.04 cm}
{\check{}})\in A$
so that the direct sum runs through all irreducible representations
of $\hat{LG}$ at level $k$.
\vs 0.2 cm

Geometric quantization of the PL model space
may be tried along the same lines. We take the complex line bundle
${\cal L}_{PL}$ over $M_{PL}$ with connection of
curvature $\Omega_{L2}$
(since the latter is complex, the connection cannot be hermitian).
Again $k\tau$ and the Cartan subgroup component of $g_0$ are
canonically conjugate and if we take a polarization annihilated by
$d\tau$, the states will be supported by $k\tau$ in the weight lattice.
For fixed $\tau$, the problem reduces to the geometric quantization on
conjugacy classes of ${\rm e}^{2\pi i\tau}$ in $G$ (the condition that
$\Omega_{L2}$ defines a Chern class of a line bundle over the
conjugacy class is exactly that $k\tau$ be a weight). For $G=SU(2)$,
the conjugacy classes are (generically) ${\bf C}P^1$ and their
(complex) symplectic form induced by $\Omega_{L2}$ corresponding to
$r$-matrix 14 is

\be
\omega_j\ =\ k(\pi i)^{-1}{\rm sin}(2\pi j/k)\hs{0.05 cm}
(|z|^2+1)^{-1}
({\rm e}^{2\pi ij/k}|z|^2+
{\rm e}^{-2\pi ij/k})^{-1}dzd\bar{z}
\en

\no for $k\tau=j\sigma^3$. $z$ is the standard complex coordinate
of ${\bf C}P^1$. Notice that $\omega_j$ is a deformation of the
symplectic
structure of a coadjoint orbit of $SU(2)$ to which it tends in the
classical
limit $k\rightarrow\infty$. For $G=SU(2)$, the quantization of the
conjugacy classes is simple since the complex structure of
${\bf C}P^1$
provides a polarization of $\omega_j$. We end up with the space
${\cal V}_{k,j}$ of holomorphic
sections of $2j$'s power of the Hopf bundle over ${\bf C}P^1$ which
may
be represented as the space of polynomials in $z$ of degree $\leq 2j$
(if we use half-densities, we should replace $k$ by $k+2$ and
take $(k+2)\tau=(j+1/2)\sigma^3$). The complete space of states
corresponding to $M_{PL}$ is then

\be
V_{PL}\ =\ \bigoplus\limits_{j=0,1/2,...,k/2}{\cal V}_{k,j}\ .
\en
\vs 0.1 cm

Usually, geometric quantization provides also prescriptions on how to
assign quantum operators to (certain) classical physical quantities.
These may be expressed in terms of a symbolic calculus using in the
case of K\"{a}hler manifolds reproducing kernels or, in more physical
terms, the formalism of coherent states$^{48}$. Although the
present case
of ${\bf C}P^1$ with form $\omega_j$ is not exactly of K\"{a}hler
type
($\omega_j$ is complex), one may set up a symbolic calculus
extending the
coherent state formalism from standard $SU(2)$ to $SU(2)_{\rm Skl}$
in such a way
that the matrix elements of $\gamma_L=g_0{\rm e}^{2\pi i\tau}
g_0^{-1}$
treated as functions of $g_0T\in{\bf C}P^1$ become generators of the
quantum deformation ${\cal U}_q(SU(2))$ of the enveloping algebra of
$SU(2)$$^{49}$ for $q={\rm e}^{\pi i/(k+2)}$. As a result,
spaces ${\cal V}_{k,j}$ carry naturally spin $j$ representations of
${\cal U}_q(SU(2))$ generated by quantizations of matrix elements of
the monodromy $\gamma_L$ satisfying the commutation
relations$^{50-52}$

\be
R^+\hs{0.06 cm}\gamma_1\hs{0.06 cm}(R^-)^{-1}\hs{0.06 cm}
\gamma_2\ =\
\gamma_2\hs{0.06 cm}R^+\hs{0.06 cm}\gamma_1\hs{0.06 cm}(R^-)^{-1}\ ,
\en

\no the quantum counterpart of Eq.\hs{0.08 cm}22.
\vs 0.2 cm

For more complicated groups than $SU(2)$, there are difficulties in
applying the standard prescriptions of geometric quantization
to the conjugacy classes of $G$ since the usual complex structure
of $G/T$ does not give a polarization for the form induced by
$\Omega_{L2}$ and there is no obvious replacement for it. It is rather
clear that some aspects of non-commutative geometry have to be used
if we want a systematic geometric procedure which produces
representation spaces ${\cal V}_{k,\lambda}$ of ${\cal U}_q(G)$
by quantizing the conjugacy classes in $G$ of
${\rm e}^{2\pi i\lambda/k}$
(or rather of ${\rm e}^{2\pi i(\lambda+\rho)/(k+h\hs{0.04 cm}
{\check{}}\hs{0.04 cm})}$\hs{0.04 cm}) taken with the symplectic
form inherited from
$\Omega_{L2}$ corresponding to $r-$matrix 14. Such a procedure
could then
be tried for non-standard solutions $r^{\pm}$ of the
CYBE and could systematically produce their quantizations $R^{\pm}$
together with the corresponding quantum
deformations of $G$ (see Ref.$^{53}$ where one of
such non-standard deformations
was analyzed). In any case, at least for standard solution 14
of the CYBE, it is reasonable to take the quantum space of states for
$M_{PL}$ as

\be
V_{PL}\ =\ \bigoplus\limits_{\lambda\ {\rm integrable}}
{\cal V}_{k,\lambda}\ .
\en

\no Then the space of quantum states which corresponds to the
fibered product
$M_{KM}\times_AM_{PL}$ clearly becomes

\be
V\ =\ \bigoplus\limits_{\lambda\ {\rm integrable}}V_{k,\lambda}\otimes
{\cal V}_{k,\lambda}\ .
\en
\vs 0.1 cm

It remains still to quantize fields $g_L(x)$. According to
Eq.\hs{0.08 cm}24,
they are built from fields $g_L^{res}(x)\equiv h(x){\rm e}^{ix\tau}$
living on $M_{KM}$ and of matrix elements of $g_0^{-1}$ defining
functions
on $M_{PL}$. $g_L^{res}(x)$ may be quantized using e.g. free field
realizations of the representations of Kac-Moody group $\hat{LG}$
(see Ref.$^{22}$ for the discussion of the $SU(2)$ case). They become
essentially chiral vertex operators of the WZW model. As for
the matrix elements of $g_0^{-1}$, we may quantize them by using
symbolic calculus for $SU(2)$ or, in general, by guesswork.
They essentially play the role of quantum group vertex operators
and may be expressed by the quantum Clebsch-Gordan coefficients.
Combination 24 of both should produce quantum field $g_L(x)$
acting in space 34 and exhibiting $G\times{\cal U}_q(G)$ symmetries
and $R$-matrix statistics:

\be
g_L(x)_1\hs{0.05 cm}g_L(y)_2\hs{0.05 cm}R^{\pm}\hs{0.05 cm}
=\hs{0.05 cm}g_L(x)_2\hs{0.05 cm}g_L(y)_1\ ,
\en

\no as discussed first in Ref.$^{10}$. It is known, however, that
the program
to construct operators
$g_L(x)$ with such properties (or its counterpart for the minimal
models)
meets difficulties due to the singular
behavior of the quantum Clebsch-Gordan coefficients at integral $k$
and has not been carried through completely yet$^
{10,13-18}$.
These difficulties seem to go back to the classical singularities
of presentation 28 of the chiral phase space $P_L$ which
breaks down for $\tau$ in Eq.\hs{0.08 cm}24 in the boundary of the
Weyl alcove.
An interesting open question is whether there exists a
quantization of
$P_L$ which does not use
separation 24 of the degrees of freedom but proceeds
directly avoiding its problems.
\vs 0.8 cm

\end{document}